\documentclass{jaa}
\usepackage{natbib}
\usepackage{url}
\usepackage{graphicx}
\usepackage{aas_macros}
\usepackage{xcolor}

\begin{document}\sloppy

\title{Moderate Nesting and Cross-Equatorial Asymmetry\\ of Active Regions in Solar Cycle 24}

\author{A.A. Norton\textsuperscript{1}, A. Mendez\textsuperscript{1}, R. Chen\textsuperscript{1}, M. Dikpati\textsuperscript{2}, S. Aswin Amirtha Raj\textsuperscript{3}}
\affilOne{\textsuperscript{1}HEPL Solar Physics, Stanford University, Stanford, CA, USA.\\}
\affilTwo{\textsuperscript{2}High Altitude Observatory, UCAR, CO, USA,\\}
\affilThree{\textsuperscript{3}Arul Anandar College, Karumathur, 625514, Madurai District, Tamil Nadu, India.}

\twocolumn[
\begin{@twocolumnfalse}

\maketitle

\corres{aanorton@stanford.edu}
\msinfo{12 October 2025}{12 October 2025}

\begin{abstract}
Solar Cycle 24 data are used to determine how often the Sun emerges sunspots in `activity nests', i.e., regions where sunspots and active regions (ARs) repeatedly emerge. We use the Solar Photospheric Ephemeral Active Region (SPEAR) catalog created from Helioseismic and Magnetic Imager (HMI) data as well as the HMI Carrington Rotation maps of radial magnetic field, $B_r$. The Sun shows moderate nesting behavior with 41\% (48\%) of AR magnetic flux found in Northern (Southern) hemispheric nests that are short-lived (average lifetimes $\sim$3.3 - 4.0 months). Different rotation rates are used to search for nests that may not be evident `by eye'. The maximum number of nests are found with slightly prograde rotational velocities, with significant nest flux also found at synodic 451--452 nHz prograde and 409--411 nHz retrograde frequencies. 
Nest patterns show strong hemispheric asymmetry, indicating that the physical origin of nests identified herein must also be asymmetric or antisymmetric across the equator.
\end{abstract}


\end{@twocolumnfalse}
]

\doinum{}
\artcitid{\#\#\#\#}
\volnum{000}
\year{0000}
\pgrange{1--}
\setcounter{page}{1}
\lp{1}

\section{Introduction}

Cool stars (0.2–1.3 $M_{\odot}$) like the Sun generate dynamic magnetic fields that influence planetary environments. These fields arise from a dynamo operating in the stellar interior that amplifies and organizes magnetic fields \citep{parker1975}. Magnetic flux emerges into the solar atmosphere in the form of sunspots, or active regions (ARs), that are magnetic in nature \citep{hale:1908} and  suppress convection and locally alter energy transport.

Although ARs have been observed for centuries, the origin and organization of the magnetic flux that produces them remain open questions. Magnetic flux emergence is widely attributed to the buoyant rise of magnetic structures through the convection zone, yet the depth and organization of the source field remain uncertain \citep{Fan2009, Cheung2010}.

\begin{figure*}[!ht]
    \centering  \includegraphics[trim=55 125 50 95, clip, width=.70\linewidth]{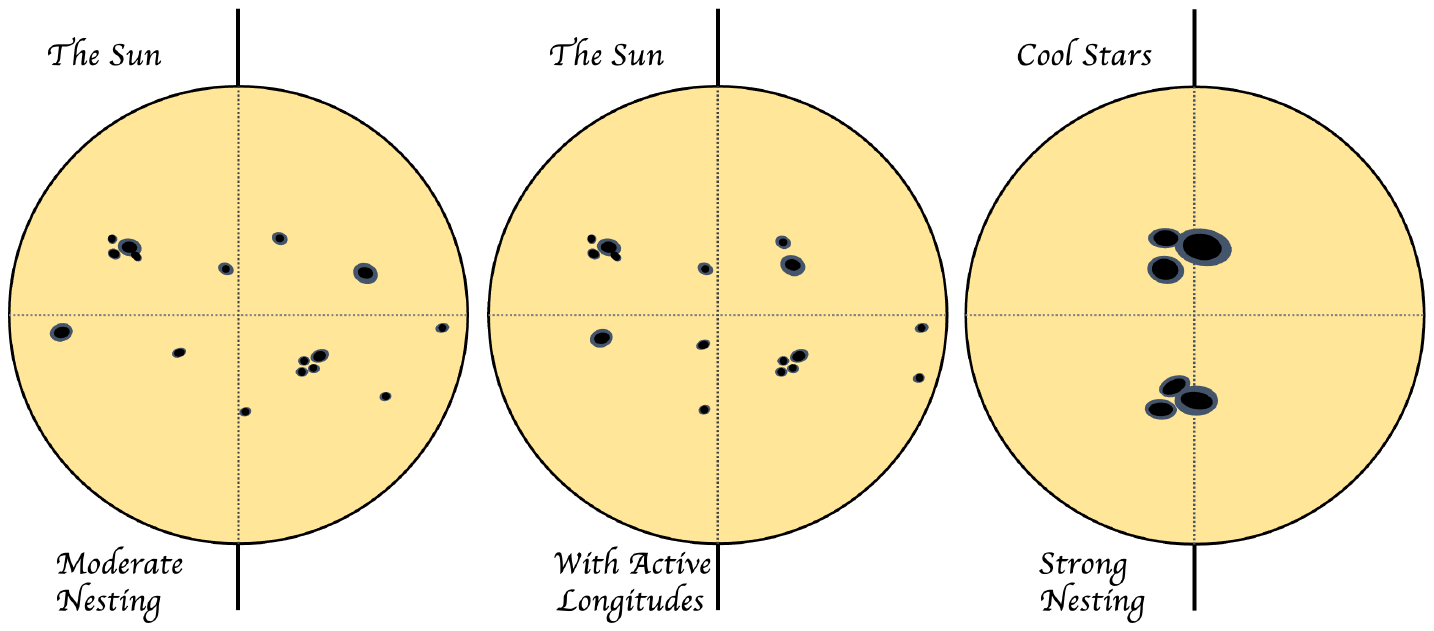}
    \caption{This cartoon depicts active region distribution similar to the Sun with a moderate degree of activity nesting (left), a moderate degree of activity nesting with active longitudes (middle), and a cool star like the Sun with stronger nesting and active longitudes. Many cool stars show a a modulation of intensity signal with rotation that is much stronger than that of the Sun.}
    \label{fig:cartoon}
\end{figure*}

A striking observational feature of ARs is their tendency to cluster in longitude and persist in preferred regions of the Sun for extended periods. These regions, often termed \textit{activity nests}, are characterized by multiple ARs emerging within a limited longitude–latitude range over several Carrington rotations. This was noted as early as 1863 by Carrington who suspected that active regions were not uniformly and randomly distributed in longitude \citep{carrington1863}. Later, \citet{bumba1969} reported on regions that emerged repeatedly with rotation rates of ~27 days and \citet{gaizauskas1983} found that these complexes of activity were maintained for 3-6 solar rotations. 

Multiple terms have been used to describe longitudinal clustering, and often the terminology is inconsistently used across the literature. Here we distinguish between related concepts: we use \textit{activity nests} to describe clustering of AR emergence within a single hemisphere, while \textit{active longitudes} refer to preferred emergence at similar longitudes in opposite hemispheres, implying symmetry across the equator (see Table~\ref{tab:table1}). 
The term \textit{hot spot} is used for regions that exhibit enhanced flare or sunspot productivity. Regardless of terminology, the longitudinal organization of solar magnetic activity provides a probe of non-axisymmetric components of the solar dynamo, which remain comparatively unexplored. Figure~\ref{fig:cartoon} depicts cartoon distributions of active regions: (left) moderate nesting similar to the Sun, (middle) moderate nesting with sunspots aligned along active longitudes, and (right) strong nesting as might be observed on a sun-like star, combined with well-defined active longitudes.

Quantitative definitions of activity nests were introduced by \citet{Castenmilleretal1986}, who identified nests as at least three ARs emerging within four Carrington rotations and within $\pm$7.5$^{\circ}$ longitude and $\pm$5$^{\circ}$ latitude, finding that 30–40\% of ARs between 1959–1964 participated in nests. The term \textit{activity nest} is often used synonymously with \textit{active longitude}. 
Some studies have described active longitudes as two preferred activity centers approximately 180$^{\circ}$ apart whose relative dominance alternates in time, a phenomenon termed as 'flip-flop' \citep{Berdyugina2003}. Other investigations use the term active longitudes more generally to denote longitudinal concentrations or recurring zones of activity without requiring that the activity centers be 180$^{\circ}$ apart \citep{Bogart1982, Bai1987, Bai2003, Henney2002}. 

%\citet{Finley2024} reported a persistent nest during Cycle~25 observed in SDO/AIA data that appeared to anchor large-scale coronal structure. 

\begin{table}[!ht]
\centering
\caption{Historical Terms for AR Clustering}
\begin{tabular}{ll}
\hline 
\hline
Term & Historical Usage\\
\hline
\textit{Activity nests} & $\geq 3$ ARs within $\pm$7.5$^{\circ}$ lon,\\
&$\pm$5$^{\circ}$ lat over $\sim$4 CRs;\\
 & Castenmiller et al. (1986).\\
\textit{Active longitudes} & Bands of recurrent ARs;  \\
& Bogart (1982), Bai (1987), \\
&Usoskin et al. (2005), \\
& Mandal (2017)\\
\textit{Hot spots} & Rigidly rotating, \\
& enhanced flaring; Bai (2003)\\
\hline
Term & Proposed Usage\\
\hline
\textit{Activity Nests} & One hemisphere only\\
& Short-lived $\leq$ 6 months\\
& Long-lived $>$ 6 months\\
\textit{Active Longitudes} 
& Nests at similar longitudes \\
& across the equator\\
\textit{Hot Spots} & Nests with flares\\
\hline
\end{tabular}
\label{tab:table1}
\end{table}

The physical origin of longitudinal clustering remains uncertain. Proposed mechanisms include modulation by giant convective cells, magneto-Rossby waves, or MHD instabilities acting on toroidal flux bands near the tachocline \citep{Zaqarashvili2010, Dikpati2018}. Flux-emergence simulations further suggest that clustering may reflect subsurface connectivity among buoyant magnetic structures \citep{Fan2009, Cheung2010}. Persistent longitudinal preferences may therefore indicate non-axisymmetric dynamo modes or interactions between global rotation and convection. Despite decades of study, it remains unclear whether longitudinal clustering reflects a persistent dynamical structure of the solar dynamo or an emergent statistical property of flux emergence.

\begin{figure*}[!ht]
    \flushleft   
    \includegraphics[trim=40 225 155 175, clip, width=0.83\textwidth]{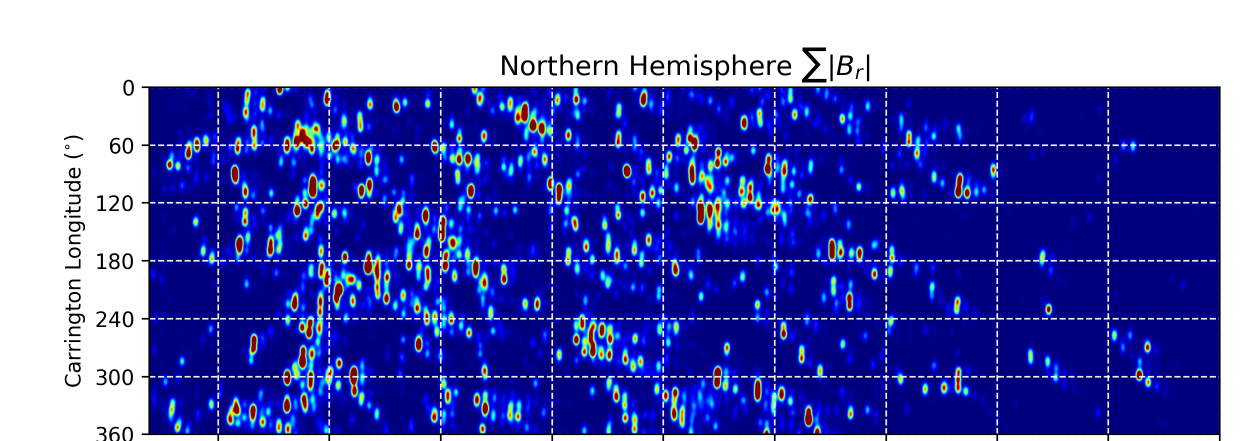}
    \includegraphics[trim=22 224 28 200, clip, width=0.84\textwidth]{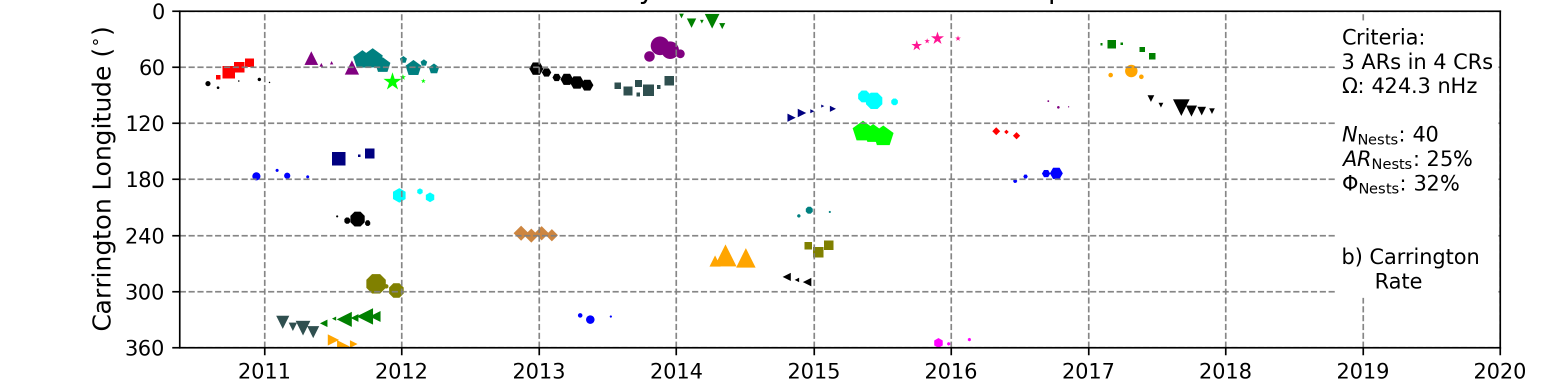}
    \includegraphics[trim=22 224 28 200, clip, width=0.84\textwidth]{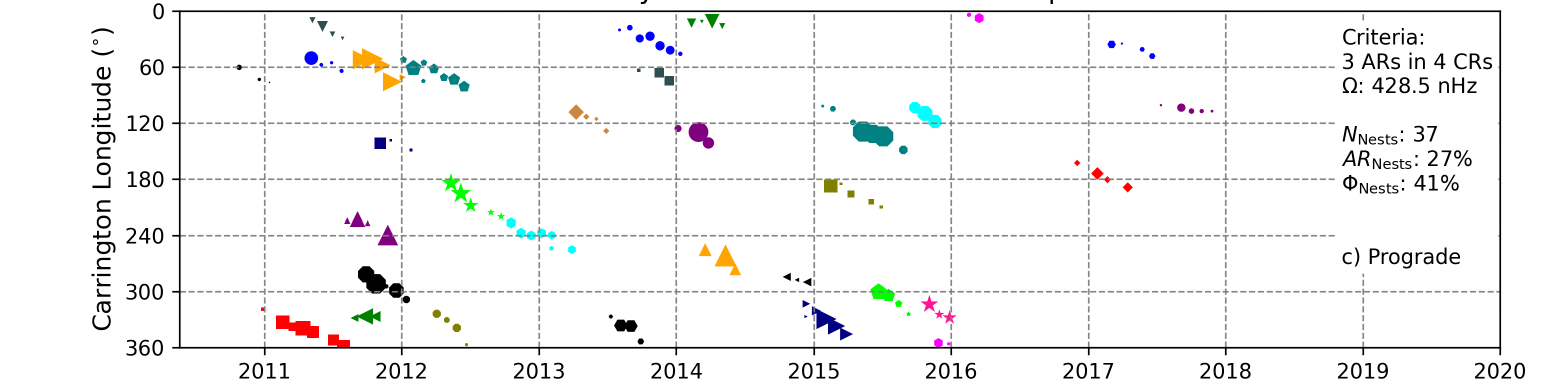}
    \caption{The North hemispheric magnetic activity is shown for Cycle 24, May 2010 \- Dec 2020, using HMI Carrington Rotation maps of $B_r$ data (top panel). The color scale is such that ARs are shown in light blue (weakest), yellow, green, and red (strongest) on a dark blue background that indicates quiet-Sun. Activity nests are identified in the SPEAR catalog data using the criteria of three ARs found within 4 CRs with $\pm$7.5$^{\circ}$ longitude and $\pm$5$^{\circ}$ latitude of each other with rotational frequency at the b) Carrington rate, and c) slightly prograde with respect to the Carrington rate. Different nests are shown using different colors while the symbol size is proportional to the amount of flux of the region. }
    \label{fig:N-nest-multi}
\end{figure*}

A central difficulty in identifying activity nests is that longitudinal organization depends on the assumed rotation frame. Most studies search for preferred longitudes in a frame rotating near the Carrington rate, implicitly assuming that longitudinal structures remain anchored in that frame \citep[e.g.,][]{Bai1987, Bai2003}. However, different rotation rates can either highlight or obscure persistent features. \citet{Bogart1982} found rotation periods differing by up to two days between solar cycles, suggesting that preferred longitudes are not rigidly tied to a single rotation rate. Other studies reported migration of active longitudes consistent with differential rotation and cycle-dependent behavior \citep{Berdyugina2003, Usoskin2005}. In this work, we expand upon previous approaches by searching across a broad range of rotation rates, allowing nests and longitudinal clustering to be identified without presupposing their anchoring frame.
%--------------------

\section{Data}

For this study, we employ synoptic Carrington rotation maps of the vector magnetic field, $B_r$, from the Helioseismic and Magnetic Imager (HMI; \citealt{Scherrer2012, Schou2012}) on board the Solar Dynamics Observatory (SDO). The synoptic maps provide full-Sun coverage of the photospheric magnetic field by assembling central meridian observations over a full Carrington rotation of 27.2753 days, enabling consistent analysis of the longitudinal distribution of solar activity. The maps can be accessed through JSOC in the data series \textit{hmi.synoptic\_mr\_polfil\_720s}. The dimensions are 3600 $\times$ 1440 with equal steps in longitude and sine latitude with values in the units of Mx cm$^{-2}$. 

In addition, we use of the Solar Photospheric Ephemeral Active Region (SPEAR) catalog \url{http://hmi.stanford.edu/hminuggets/?p=3730}, that is based on the Space-weather HMI Active Region Patch (SHARP) data product. SHARPs identify and track ARs in near-real time using HMI vector magnetograms \citep{Bobra2014}, and the SPEAR catalog provides a tabulated set of data by compiling AR properties such as total flux, latitude and longitude location, tilt angle, etc., when the SHARP regions are nearest to the central meridian. Each SHARP region is only found once in the catalog. 

The monthly hemispheric sunspot numbers provided by the World Data Center for the Sunspot Index and Long-term Solar Observations (WDC-SILSO) are used to show the evolution of Solar Cycle 24 in the North and South hemispheres \citep{SILSO2026}. Monthly values are calculated as averages of daily hemispheric counts and are distributed in the Version 2.0 SILSO data release, which provides a recalibrated and internally consistent extension of the international sunspot number series. The dataset, available since 1992, can be found here; \url{https://www.sidc.be/SILSO/datafiles}.

\section{Methods}

\begin{figure*}[!ht] 
    \includegraphics[trim=40 225 155 175, clip, width=0.83\textwidth]{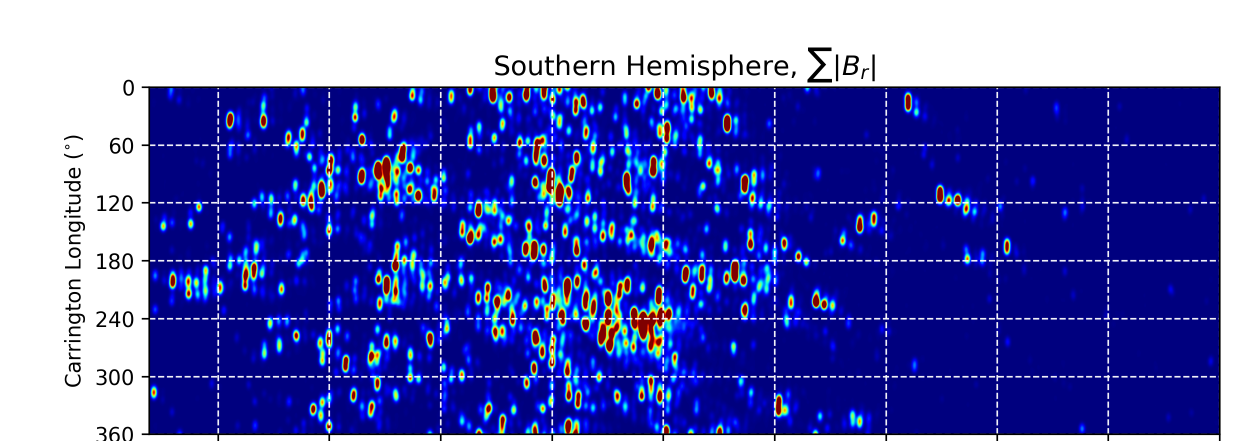}
    \includegraphics[trim=22 224 28 200, clip, width=0.84\textwidth]{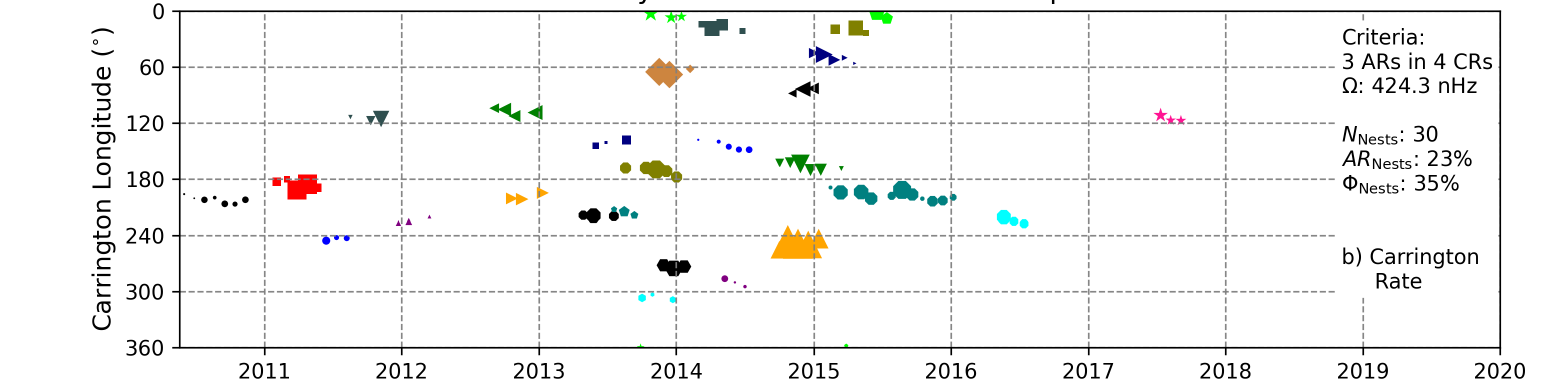}
    \includegraphics[trim=22 224 28 200, clip, width=0.84\textwidth]{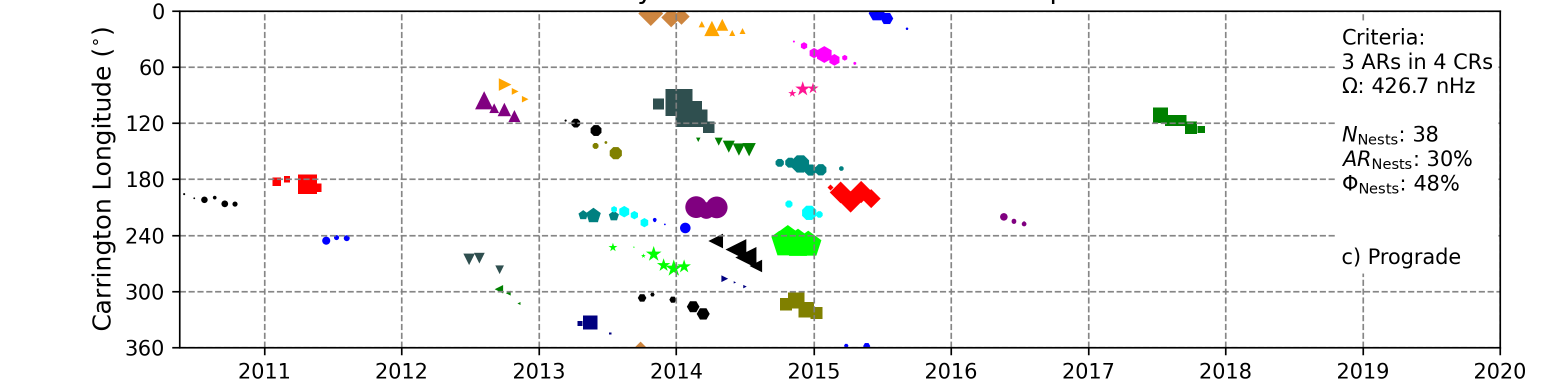}
    \caption{The same as Figure 2 but for the Southern hemisphere. }
    \label{fig:S-nest-multi}
\end{figure*}

We use HMI Carrington Rotation maps of the radial vector magnetic field, $B_r$, for visualization and comparison purposes only (not quantitative identification of nests), in order to see the times and locations of strong active region magnetic flux and compare the sunspot number with the radial magnetic flux time series from HMI.  We plot the sum of the absolute value of $B_r$ between the latitudes of $0-30^{\circ}$, in the top panel of Figures \ref{fig:N-nest-multi} and \ref{fig:S-nest-multi}, keeping the hemispheres separate in order to note the northern and southern hemisphere patterns. We downsample the Carrington longitude from 3600 to 360 pixels and use a 2-dimensional Gaussian smoothing with a FWHM of 7$^{\circ}$ in longitude and 2 Carrington rotations in time. Note that the quantitative analysis to identify nests uses data in the SPEAR catalog, and does not have any smoothing. Also, since the Sun rotates differentially with higher latitudes rotating slower than those close to the equator, the Carrington rate is used to build the synoptic maps but is not necessarily indicative of the rate at which magnetic features rotate.

We used the SPEAR catalog to identify nests. ARs are recorded in the SPEAR catalog only once when they are closest to the central meridian. We remove long-lived ARs that return to the front side of the Sun for a second or third time. We remove returned ARs by first using a method developed by \citet{Yeates:2020} using flux transport models to estimate what a given AR rotating off the front side of the disk would look like upon its return. If the forward-modeled AR is a good match for the observed AR, then the observed AR NOAA number and SHARP number is listed as a returning sunspot in a separate file.  The long-lived and returned ARs are then not allowed to be a member of any nest. 

To identify a nest, each AR is used as a starting data point, then nearby latitudes ($\pm$7.5$^{\circ}$), longitudes ($\pm$5$^{\circ}$), and Carrington Rotations ($\pm$4 Carrington Rotations) are searched to find other ARs that could be part of the nest given a certain rotation rate and nest definition. If a rotation rate is specified that is different from the Carrington rate of 424.3427 nHz, then the search for ARs in time updates the search locations to be $\pm\Delta$longitude corresponding to the change in longitude per rotation that would correspond to that rotation rate. If an AR is identified in two nests, then those nests are merged. All frequencies in this paper are synodic, not sidereal. To convert to sidereal rates, add 31.6875 nHz to the synodic rate.

To test the significance of our results in identifying nests, the longitudes of the regions in the SPEAR catalog are randomized for 10,000 trials and nests are identified for each trial.  The time and latitude information of the ARs is not randomized. We record the number of ARs and the percent of flux found in nests within the randomized trials.  The one, two and three sigma levels are then calculated for the 10,000 trials, assuming the random data contain no nests. Observational results with a percent flux value in nests above the two or three sigma levels in the randomized trials are noted. Nests are identified for a range of rotation rates.

In order to test the hemispheric asymmetry of ARs and nests, we first determine what fraction of the time an AR or nest in one hemisphere will have an AR or nest  at the same longitude in the other hemisphere. We first plot the locations of all ARs from each hemisphere as a function of Carrington rotation and longitude and calculate how often they intersect, or occupy, the same longitude. This is the observed likelihood that an AR in one hemisphere will have an AR in a similar location in the other hemisphere. We then do the same for the identified nests. We report on how likely it is for a nest in one hemisphere to exist in the same longitude and time bin in the other hemisphere, simply from the intersection percentage. To test how likely the intersection of N-S nests are in a randomized distribution, the central longitude of each nest is randomized but the relative location of each AR within a given nest with respect to the central longitude is kept constant. This randomizes the nest locations in longitude but ensures the nest retains the same number of ARs. The time of each nest is not randomized for this test. 

\section{Results}

The amount of nesting exhibited by the Sun depends on the definition used to describe what constitutes an activity nest. In this manuscript, we search for nests using the \citet{Castenmilleretal1986} definition with variable rotation rates. 

\begin{figure*}[!ht]
    \flushleft   
    \includegraphics[trim=0 160 40 125, clip, width=0.80\textwidth]{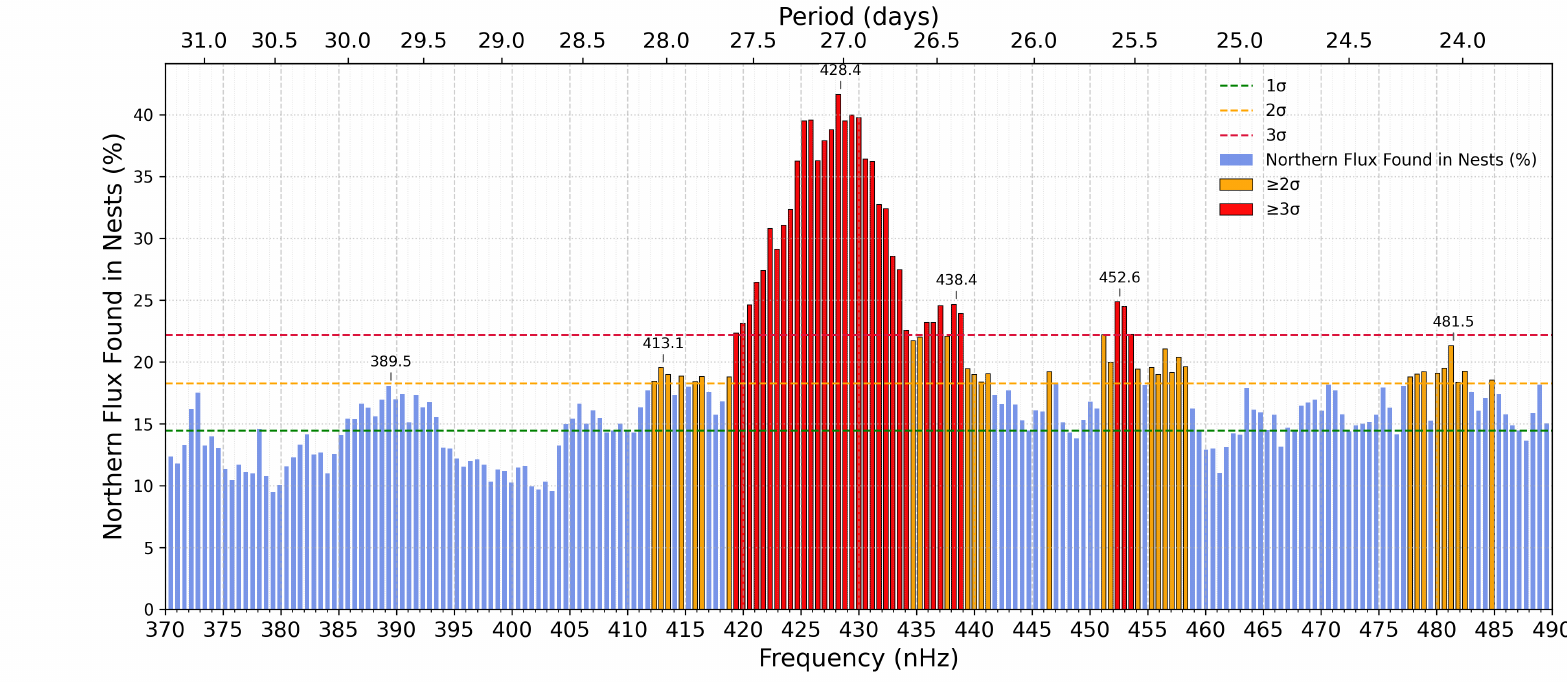}
    \includegraphics[trim=0 160 40 125, clip, width=0.80\textwidth]{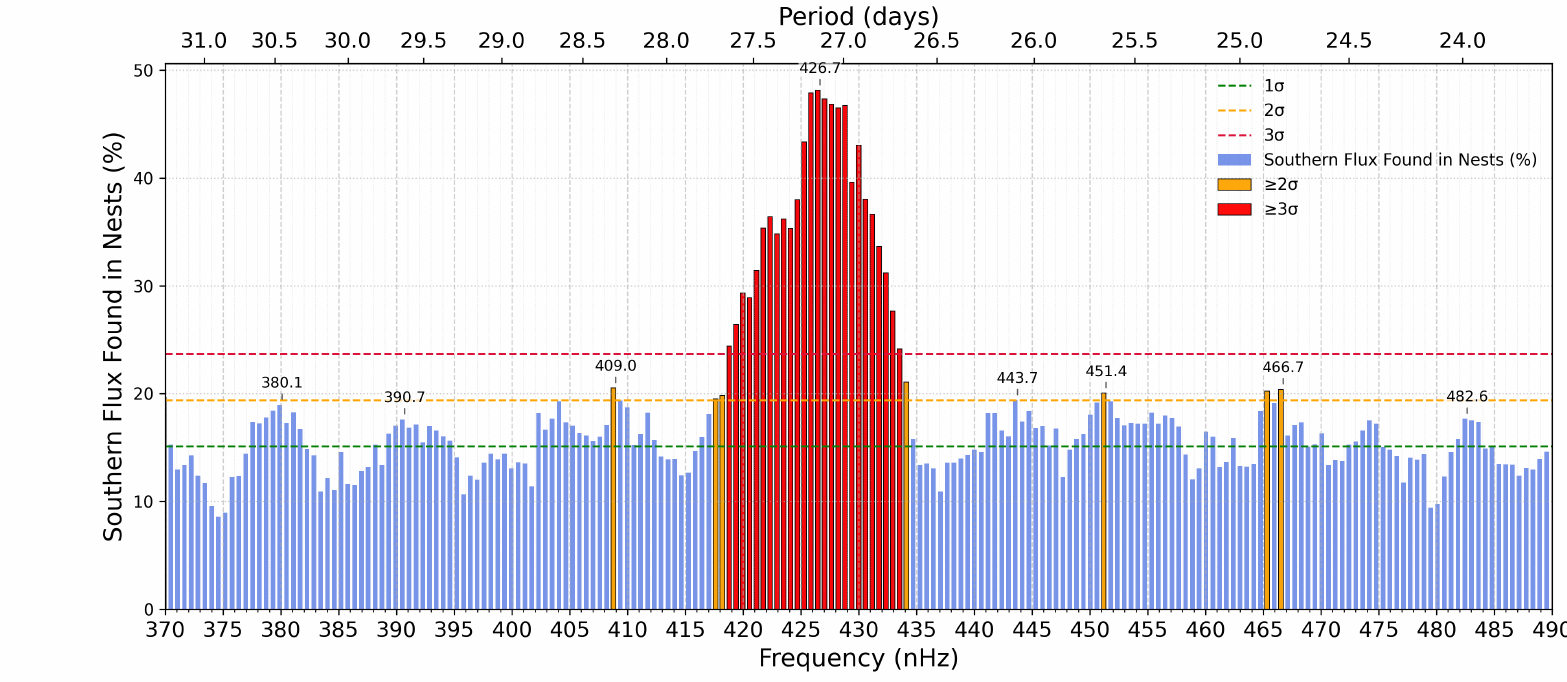}
    \caption{The percent of flux found in nests with criteria of three ARs within four CRs with $\pm$7.5$^{\circ}$ longitude and $\pm$5$^{\circ}$ latitude of each other is shown as a function of rotational frequency. One, two and three sigma statistical significance levels are plotted as horizontal lines determined from 10,000 trials where the longitudes of the ARs are randomized.  Note that other quantities associated with the AR - latitude, time and flux - are not randomized. Rotational frequencies are shown in red (orange) for frequencies if the amount of flux found in nests is above 3 (2)-$\sigma$ significance levels. }
    \label{fig:rates-pvalues1}
\end{figure*}
\subsection{Degree of Nesting}

We first examine the Carrington rotation synoptic maps of $B_r$ for the HMI data from Solar Cycle 24, shown in the top panels of Figures \ref{fig:N-nest-multi} and \ref{fig:S-nest-multi} in which strong magnetic flux patterns can be observed as a function of Carrington latitude versus time. Patterns that form a line sloping downward and to the right in the figure indicate prograde features, i.e. meaning the pattern is moving faster that the Carrington rotation rate, while patterns that slope upward and to the right are indicative of retrograde motion.  

Using the criteria of \citet{Castenmilleretal1986}, in which the nest critera is three ARs found within four CRs with $\pm$7.5$^{\circ}$ longitude and $\pm$5$^{\circ}$ latitude of each other, the SPEAR catalog is used to search through the AR locations to find clustering. When using the Carrington rate to search the catalog, 40 nests with 25\% of ARs and 32\% of flux are identified in the North and 30 nests with 23\% of ARs and 35\% of flux are found in the South, see the middle panel in Figures \ref{fig:N-nest-multi} and \ref{fig:S-nest-multi} that shows separate nests plotted as different symbols and colors. Note that latitude is not shown in the figures so that even if there is a large amount of flux in a location, a nest might not be found if the ARs are further away from each than $\pm$5$^{\circ}$ latitude.

If we used $\pm$6$^{\circ}$ in latitude to search for nests, the amount of flux in nests increases 2\% in both the North and South hemispheres. If using $\pm$7$^{\circ}$  in latitude, it increases another 2-3\%, continuing in this trend until it saturates at $\pm$15$^{\circ}$ latitude (the full width of the active latitudes) at the values of 43\% (49\%) flux found in nests in the N (S) hemisphere. The $\pm$5$^{\circ}$ in latitude represents a distance of $\sim$120 Mm, meaning ARs can be 120 Mm apart in latitude and still be in the same nest using this definition. Given that supergranules are 30 Mm in diameter, this would be a nest spanning four supergranules in latitude which is an appropriate, albeit large, distance for a nest in latitude.  If we were searching for active longitudes, we would not limit the latitudinal extent at all. 

The number of ARs and AR flux in nests increases when the data is searched using slightly prograde rates of rotation. The maximum amount of AR flux participating in nests is 48\% (41\%) in the South (North) hemisphere at rotational rates of 428.5 (426.7) nHz, slightly prograde with respect to the Carrington rate, see bottom panel in Figures \ref{fig:N-nest-multi} and Figure \ref{fig:S-nest-multi}.  These are considered short-lived nests since they live, on average, between 3.3$-$4 Carrington Rotations.  

%Note: Lifetimes 3.3 CR N and 3.4 CR Swith 424.3 
%428.5  3.9 CR N and 3.3 CR S with 428.5
%3.5 CRs for S with 326.7

\begin{figure*}[!ht]
    \flushleft   
    \includegraphics[angle=-90, trim=210 50 210 50, clip, width=0.85\textwidth]{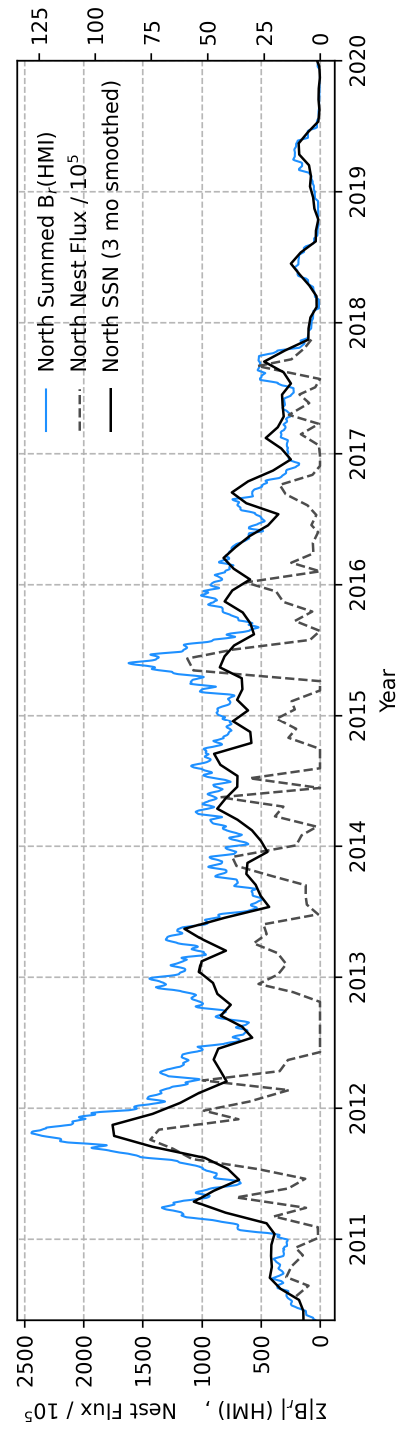}
    \includegraphics[angle=-90, trim=210 50 210 50, clip, width=0.85\textwidth]{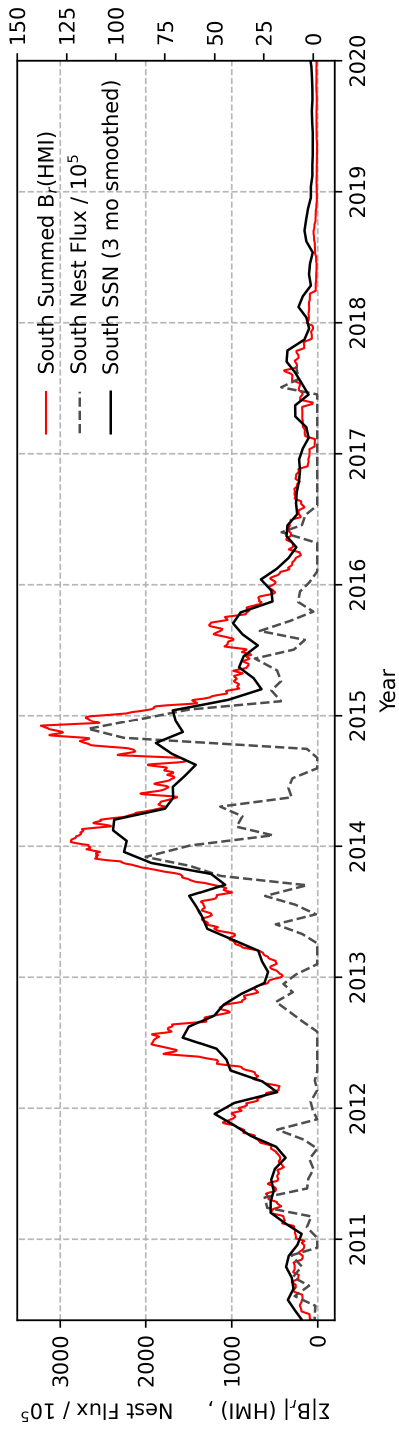}
    \includegraphics[angle=-90,trim=190 10 200 50, clip, width=0.825\textwidth]{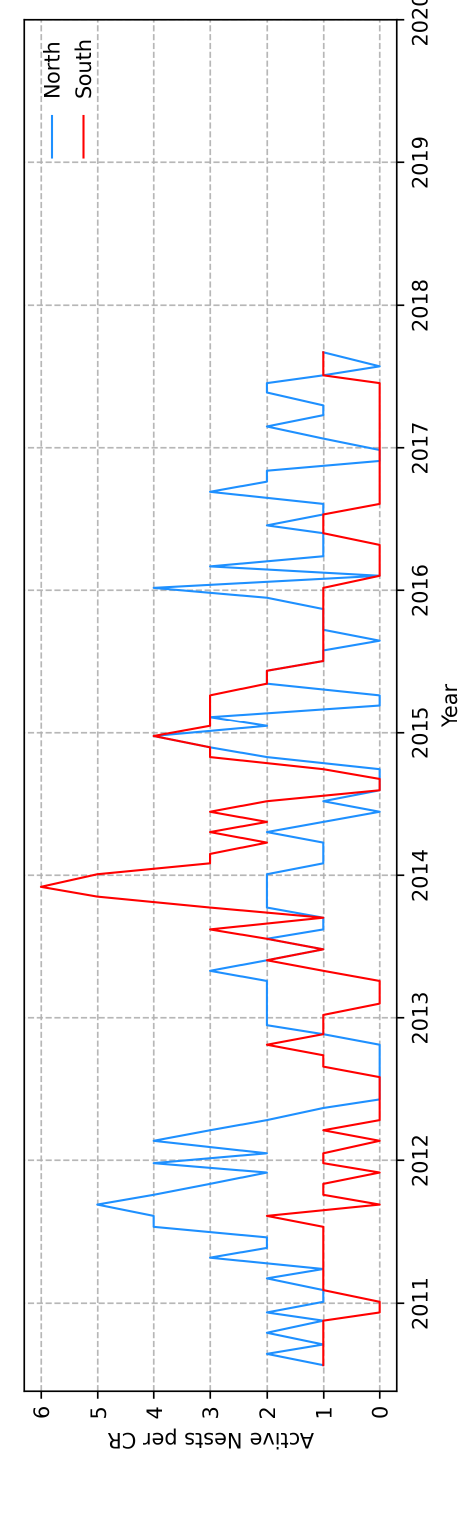}
    \caption{The amount of magnetic flux in activity nests (gray dashed line, Castenmiller criteria and Carrington rotation rate) is shown in context with the monthly hemispheric sunspot number (black line) as a function of time for the North (top panel) and South hemisphere (middle panel). The plotted sunspot number has a boxcar smoothing of 3 months. Overplotted is a time series of HMI B$_r$ values summed over the North (blue line) and South (red line) latitudes of 0-30$^{\circ}$ in a width of 1$^{\circ}$ longitude increments in the Carrington rotation data, which allows for a time series to be created from the synoptic data. The plotted B$_r$ data has a boxcar smoothing of 3 months. The number of activity nests in the North (blue) and South hemispheres (red line) as a function of time are shown in the bottom panel.}
    \label{fig:Cyclevar}
\end{figure*}

\subsection{Nests at Different Rotation Rates}

The percent of flux found in nests as a function of rotation rate is shown in Figure \ref{fig:rates-pvalues1} for the \citet{Castenmilleretal1986} nest criteria, i.e. 3 ARs in 4 CRs, basically short-lived nests.  The North and South hemisphere show significant flux in nests for rates near the Carrington rotation rate, i.e. synodic frequencies of 418-433 nHz synodic, or 27.7 -26.7 days. Other frequencies that contain significant flux in nests are found at 451-452 nHz and 409-413 nHz. These may be related to other processes such as inertial modes. 

\subsection{Nest Flux and Number During the Solar Cycle}
The temporal evolution of magnetic flux contained within activity nests broadly follows the hemispheric solar cycle progression as traced by the monthly hemispheric sunspot number, see Figure \ref{fig:Cyclevar}. In both hemispheres, nest flux peaks near the epochs of maximum activity. Over the interval May 2010–December 2019, the total magnetic flux contained in nests is comparable between hemispheres, with $1.48\times10^{9}$ G in the North and $1.63\times10^{9}$ G in the South, corresponding to a North/South ratio of 0.91.

Despite the overall correlation with sunspot number, the relationship is not strictly proportional. For example, during mid-2012 in the Southern hemisphere the sunspot number exhibits a pronounced enhancement that is not accompanied by a corresponding increase in nest flux, indicating that elevated activity levels do not always manifest as strongly clustered emergence. The number of simultaneously present activity nests varies throughout the cycle, reaching maxima of five nests in the North and six in the South. In each hemisphere, the largest number of nests occurs close to the epoch of peak sunspot activity which is late 2011 for the North and late 2013 to early 2014 for the South, suggesting that nesting becomes most prevalent near cycle maximum.

\subsection{Hemispheric Asymmetry}
\begin{figure*}[!ht]
    \flushleft   
    \includegraphics[trim=70 190 100 170, clip, width=0.63\textwidth]{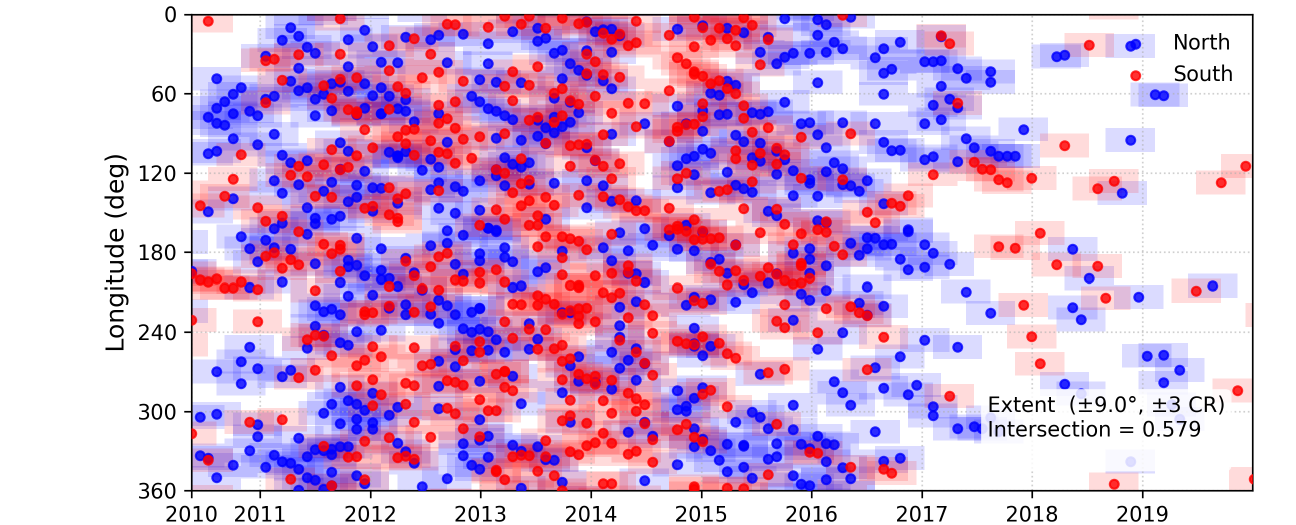}
    \includegraphics[trim=70 190 100 160, clip, width=0.63\textwidth]{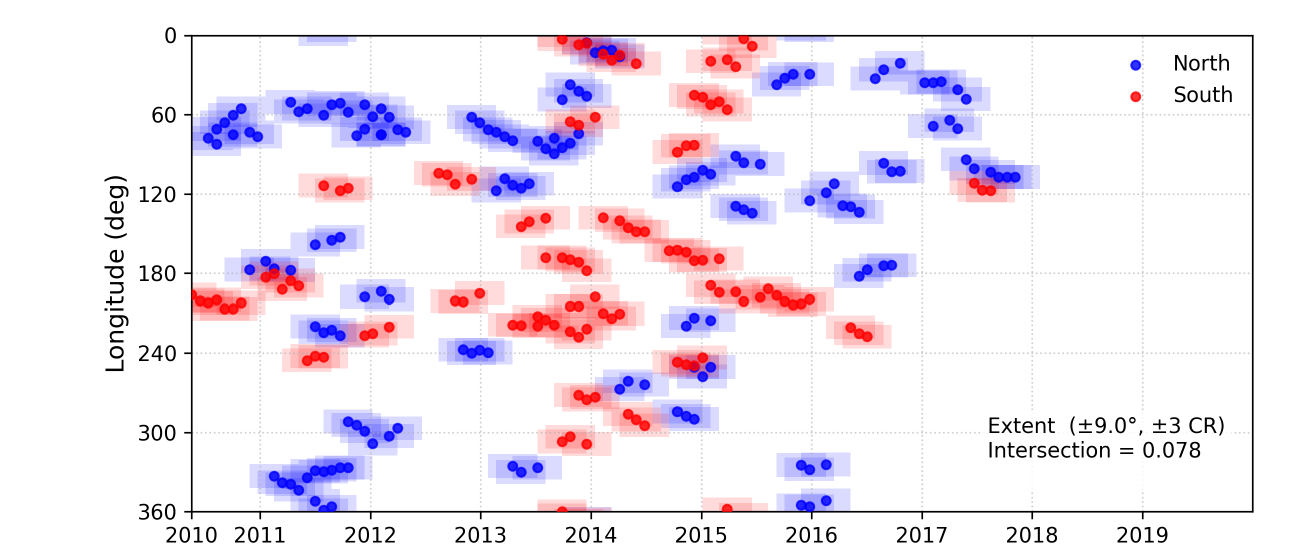}
    \includegraphics[trim=55 170 100 160, clip, width=0.625\textwidth]{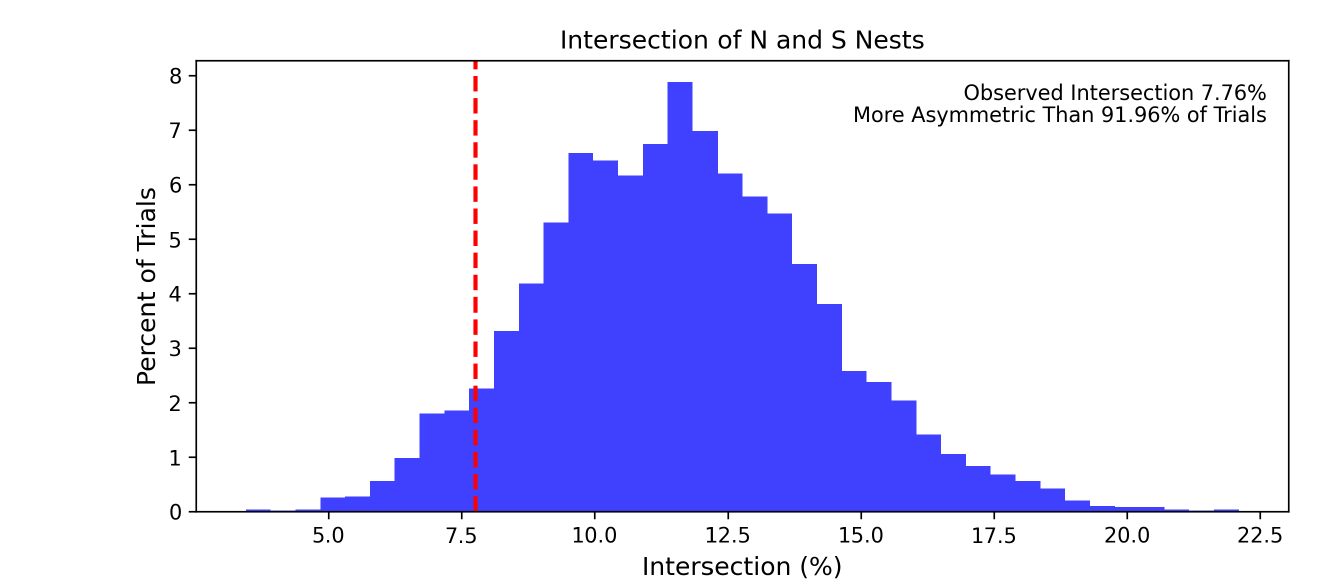} 
    \caption{The distribution of individual ARs from the North and South hemispheres is shown (top panel) in blue and red symbols with some dilation of $\pm$9$^{\circ}$ longitude and $\pm$3 Carrington rotations. The number of intersections of individual ARs in the North and South is noted as 0.579 ($\sim$58\%). Nests identified using the Castenmiller criteria of 3 ARs in 4 CRs are shown (middle panel) with only 7.8\% of the ARs in nests intersecting. Randomization trials show that this hemispheric asymmetry in the nest locations is more asymmetric than $\sim$92\% of trials.}
    \label{fig:asym}
\end{figure*}

We examine the locations of all ARs that are assigned a NOAA number and also have an area greater than 50 $\mu$Hemispheres. Separating the ARs into North and South hemispheres, we plot them as a function of Carrington Rotation and Carrington Longitude, see Figure \ref{fig:asym}, top panel. After dilating their locations by $\pm9^{\circ}$ longitude and $\pm$3 Carrington rotations to provide some width, we count the percentage of ARs that intersect across the Northern and Southern hemisphere. The fraction of ARs that overlap in North and South bins with this dilation is $\sim$58\%. This can be interpreted as follows: given an AR in one hemisphere, there is a 58\% chance that an AR will be located within $\pm$9$^{\circ}$ longitude and $\pm$3 Carrington rotations in the opposite hemisphere. 

The same is done for the nest population of ARs, see the middle panel of Figure \ref{fig:asym}, using the Castenmiller criteria with a Carrington rotation rate. Of all the longitude-time bins occupied by an AR in a nest in one hemisphere, only $\sim$7.8\% have an AR in a nest in the opposite hemisphere within $\pm$3 CR in time and $\pm9^{\circ}$ in longitude. This means that when nests form in one hemisphere, they are overwhelmingly absent at the corresponding longitude and time in the opposite hemisphere. In contrast to the broader active region population, which shows substantial co-occupancy, the organization of nests is characterized by strong hemispheric antisymmetry. Identifying nests using a slightly prograde rotation rate (the rates shown in Figures 2 and 3 when the maximum amount of flux is found in nests) provides a North-South ARs in nests intersection of 11.9\%. 

We determine the significance of the nest asymmetry by randomizing the central longitude of the nest locations. Meaning, we characterize the asymmetry by randomly shifting the average longitude of each nest while keeping the number of ARs in the nest constant. We do not randomize the latitude or time of the nests. The 10,000 trials show that the observed nest asymmetry is greater than 92\% of the randomized trials, see Figure \ref{fig:asym} lower panel. If using the slightly prograde rotation rates to find nests, the nest asymmetry is greater than 83.4\% of trials.

\section{Conclusions}

The two most significant findings of this work are that 1) 40 - 50\% of AR flux is found in short-lived nests, and 2) the nests are asymmetric across the equator indicating that the origins of nests are not giant convection cells that span the equator with hemispheric symmetry. 

Although the raw distribution of individual ARs appears predominantly symmetric — with $\sim$58\% of longitude–time bins co-occupied by an AR in both hemispheres — Monte Carlo tests of 10,000 randomized trials show that the the observed configuration of nests is more asymmetric than 92\% of the null cases. \citet{dikpati2003} found that antisymmetric modes dominate (see their Figure 6) for the dynamo-generated, solar-like toroidal bands at or below $15^{\circ}$ latitude, i.e., when the solar activity peaks and/or starts declining. This theoretical finding may be consistent with the observed antisymmetric-type distribution of active regions.  

It may be that an antisymmetric mode, an inertial mode for example, is influencing nesting in which case one might expect nests in one hemisphere to be prograde and in the other hemisphere to be retrograde for a certain amount of time.  We have not searched the parameter space for antisymmetry in this way. 

The number of nests and amount of flux in nests tracks the solar cycle evolution but does not simply scale with overall activity level. While both the magnetic flux contained in nests and the number of nests peak near hemispheric sunspot maxima, periods of enhanced sunspot number can occur without a corresponding increase in nesting. The maximum number of nests at a given time were five in the North and six in the south. The ratio of the total amount of flux in the North and South hemispheres was 0.91. 

Note that solar rotation rates have been reported to depend on the magnetic activity of the Sun \citep{hathaway1990, wang2022}, with faster rotation rates occurring during less magnetically active cycles and different rotation rates found in the hemispheres due to varying levels of magnetism during a given cycle. 

In a future publication, we intend to report on the prevalence of long-lived nests (lifetimes $>$ 6 months), longitudinal modes ($m=1, 2, 4$, etc.), and the results from wavelet analysis. Deeper analysis of the asymmetry is warranted as it suggests a way we can ascertain the mechanism influencing nesting. 

\section*{Acknowledgements}
This work was supported by NASA HSR grants NNH18ZDA001N and NNH21ZDA001N, and NASA DRIVE Center COFFIES grant 80NSSC20K0602.

\vspace{-1em}

\bibliography{nests}{}

@ARTICLE{Bogart1982,
       author = {{Bogart}, R.~S.},
        title = "{Recurrence of solar activity}",
      journal = {\solphys},
         year = 1982,
        month = sep,
       volume = {76},
        pages = {155-165},
          doi = {10.1007/BF00170987},
       adsurl = {https://ui.adsabs.harvard.edu/abs/1982SoPh...76..155B}
}

@ARTICLE{parker1975,
       author = {{Parker}, E.~N.},
        title = "{The generation of magnetic fields in astrophysical bodies. X. Magnetic buoyancy and the solar dynamo.}",
      journal = {\apj},
         year = 1975,
        month = may,
       volume = {198},
        pages = {205-209},
          doi = {10.1086/153593},
       adsurl = {https://ui.adsabs.harvard.edu/abs/1975ApJ...198..205P}
}

@ARTICLE{Bai1987,
       author = {{Bai}, T.},
        title = "{Distribution of flares on the sun during solar cycle 19-23: Active longitudes and the periodicity}",
      journal = {\apj},
         year = 1987,
        month = aug,
       volume = {318},
        pages = {L63-L67},
          doi = {10.1086/184947},
       adsurl = {https://ui.adsabs.harvard.edu/abs/1987ApJ...318L..63B}
}

@ARTICLE{Bai2003,
       author = {{Bai}, T.},
        title = "{Hot Spots for Solar Flares Persisting for Decades: Longitude Distributions of Flares of Cycles 19-23}",
      journal = {\apj},
         year = 2003,
        month = jul,
       volume = {585},
       number = {2},
        pages = {1114-1123},
          doi = {10.1086/346151},
       adsurl = {https://ui.adsabs.harvard.edu/abs/2003ApJ...585.1114B}
}

@ARTICLE{Henney2002,
       author = {{Henney}, C.~J. and {Harvey}, J.~W.},
        title = "{The Global Solar Magnetic Field Distribution over Three Solar Cycles}",
      journal = {\solphys},
         year = 2002,
        month = jan,
       volume = {207},
        pages = {199-217},
          doi = {10.1023/A:1014913700206},
       adsurl = {https://ui.adsabs.harvard.edu/abs/2002SoPh..207..199H}
}

@ARTICLE{Zaqarashvili2010,
       author = {{Zaqarashvili}, T.~V. and {Carbonell}, M. and {Oliver}, R. and {Ballester}, J.~L.},
        title = "{Rossby waves in the solar tachocline and Rieger-type periodicities}",
      journal = {\apj},
         year = 2010,
        month = apr,
       volume = {709},
       number = {2},
        pages = {749-758},
          doi = {10.1088/0004-637X/709/2/749},
archivePrefix = {arXiv},
       eprint = {0912.4593},
 primaryClass = {astro-ph.SR},
       adsurl = {https://ui.adsabs.harvard.edu/abs/2010ApJ...709..749Z}
}

@ARTICLE{dikpati2003,
       author = {{Dikpati}, Mausumi and {Gilman}, Peter A. and {Rempel}, Matthias},
        title = "{Stability Analysis of Tachocline Latitudinal Differential Rotation and Coexisting Toroidal Band Using a Shallow-Water Model}",
      journal = {\apj},
         year = 2003,
        month = oct,
       volume = {596},
       number = {1},
        pages = {680-697},
          doi = {10.1086/377708},
       adsurl = {https://ui.adsabs.harvard.edu/abs/2003ApJ...596..680D}
}

@ARTICLE{Dikpati2018,
       author = {{Dikpati}, M. and {Belucz}, B. and {Gilman}, P.~A.},
        title = "{A Shallow-water Model for the Solar Tachocline II: Magneto-shear Instabilities}",
      journal = {\apj},
         year = 2018,
        month = jul,
       volume = {857},
       number = {2},
        pages = {55},
          doi = {10.3847/1538-4357/aab7f7},
archivePrefix = {arXiv},
       eprint = {1803.01962},
 primaryClass = {astro-ph.SR},
       adsurl = {https://ui.adsabs.harvard.edu/abs/2018ApJ...857...55D}
}

@ARTICLE{Berdyugina2003,
       author = {{Berdyugina}, S.~V. and {Usoskin}, I.~G.},
        title = "{Active longitudes in sunspot activity: Century scale persistence and north-south asymmetry}",
      journal = {\aap},
         year = 2003,
        month = mar,
       volume = {405},
        pages = {1121-1128},
          doi = {10.1051/0004-6361:20030748},
       adsurl = {https://ui.adsabs.harvard.edu/abs/2003A&A...405.1121B}
}

@ARTICLE{Usoskin2005,
       author = {{Usoskin}, I.~G. and {Berdyugina}, S.~V. and {Poutanen}, J.},
        title = "{Preferred longitudes of sunspot activity: Evidence for active longitudes}",
      journal = {\aap},
         year = 2005,
        month = oct,
       volume = {441},
        pages = {347-352},
          doi = {10.1051/0004-6361:20053311},
archivePrefix = {arXiv},
       eprint = {astro-ph/0508460},
 primaryClass = {astro-ph},
       adsurl = {https://ui.adsabs.harvard.edu/abs/2005A&A...441..347U}
}

@ARTICLE{Scherrer2012,
       author = {{Scherrer}, P.~H. and {Schou}, J. and {Bush}, R.~I. and {Kosovichev}, A.~G. and {Bogart}, R.~S. and {Hoeksema}, J.~T. and {Liu}, Y. and {Duvall}, T.~L. and {Zhao}, J. and {Title}, A.~M. and {Schrijver}, C.~J. and {Tarbell}, T.~D. and {Tomczyk}, S.},
        title = "{The Helioseismic and Magnetic Imager (HMI) Investigation for the Solar Dynamics Observatory (SDO)}",
      journal = {\solphys},
         year = 2012,
        month = jan,
       volume = {275},
       number = {1-2},
        pages = {207-227},
          doi = {10.1007/s11207-011-9834-2},
       adsurl = {https://ui.adsabs.harvard.edu/abs/2012SoPh..275..207S}
}

@ARTICLE{Schou2012,
       author = {{Schou}, J. and {Scherrer}, P.~H. and {Bush}, R.~I. and {Wachter}, R. and {Couvidat}, S. and {Rabello-Soares}, M.~C. and {Bogart}, R.~S. and {Hoeksema}, J.~T. and {Liu}, Y. and {Duvall}, T.~L. and {Akin}, D.~J. and {Allard}, B.~A. and {Miles}, J.~W. and {Rairden}, R. and {Shine}, R.~A. and {Tarbell}, T.~D. and {Title}, A.~M. and {Wolfson}, C.~J. and {Elmore}, D.~F. and {Norton}, A.~A. and {Tomczyk}, S.},
        title = "{Design and Ground Calibration of the Helioseismic and Magnetic Imager (HMI) Instrument on the Solar Dynamics Observatory (SDO)}",
      journal = {\solphys},
         year = 2012,
        month = jan,
       volume = {275},
       number = {1-2},
        pages = {229-259},
          doi = {10.1007/s11207-011-9842-2},
       adsurl = {https://ui.adsabs.harvard.edu/abs/2012SoPh..275..229S}
}

@ARTICLE{Bobra2014,
       author = {{Bobra}, M.~G. and {Sun}, X. and {Hoeksema}, J.~T. and {Turmon}, M. and {Liu}, Y. and {Hayashi}, K. and {Barnes}, G. and {Leka}, K.~D.},
        title = "{The Helioseismic and Magnetic Imager (HMI) Vector Magnetic Field Pipeline: SHARPs - Space-weather HMI Active Region Patches}",
      journal = {\solphys},
         year = 2014,
        month = nov,
       volume = {289},
       number = {9},
        pages = {3549-3578},
          doi = {10.1007/s11207-014-0529-3},
       adsurl = {https://ui.adsabs.harvard.edu/abs/2014SoPh..289.3549B}
}

@book{carrington1863,
  author    = {Carrington, Richard Christopher},
  title     = {Observations of the Spots on the Sun from November 9, 1853, to March 24, 1861, Made at Redhill},
  year      = {1863},
  publisher = {Williams and Norgate},
  address   = {London},
  url       = {https://archive.org/details/observationsofsp00carr}
}

@ARTICLE{hathaway1990,
       author = {{Hathaway}, David H. and {Wilson}, Robert M.},
        title = "{Solar Rotation and the Sunspot Cycle}",
      journal = {\apj},
         year = 1990,
        month = jul,
       volume = {357},
        pages = {271},
          doi = {10.1086/168913},
       adsurl = {https://ui.adsabs.harvard.edu/abs/1990ApJ...357..271H}
}

@ARTICLE{wang2022,
       author = {{Wang}, J. and {Wang}, J. and {Wang}, L.~L. and {Sun}, W. and {Xiao}, Z.~Y. and {Zhang}, H. and {Liang}, Z.},
        title = "{Analysis of the Relationship between Solar Activity and Solar Rotation Rate}",
      journal = {Acta Astronomica Sinica},
         year = 2022,
        month = may,
       volume = {63},
       number = {3},
          eid = {34},
        pages = {34},
       adsurl = {https://ui.adsabs.harvard.edu/abs/2022AcASn..63...34W}
}

@ARTICLE{Yeates:2020,
       author = {{Yeates}, Anthony R.},
        title = "{The Minimal Helicity of Solar Coronal Magnetic Fields}",
      journal = {\apjl},
         year = 2020,
        month = aug,
       volume = {898},
       number = {2},
          eid = {L49},
        pages = {L49},
          doi = {10.3847/2041-8213/aba762},
archivePrefix = {arXiv},
       eprint = {2007.10649},
 primaryClass = {astro-ph.SR},
       adsurl = {https://ui.adsabs.harvard.edu/abs/2020ApJ...898L..49Y}
}

@ARTICLE{hale:1908,
       author = {{Hale}, George E.},
        title = "{On the Probable Existence of a Magnetic Field in Sun-Spots}",
      journal = {\apj},
         year = 1908,
        month = nov,
       volume = {28},
        pages = {315},
          doi = {10.1086/141602},
       adsurl = {https://ui.adsabs.harvard.edu/abs/1908ApJ....28..315H}
}

@ARTICLE{gaizauskas1983,
       author = {{Gaizauskas}, V. and {Harvey}, K.~L. and {Harvey}, J.~W. and {Zwaan}, C.},
        title = "{Large-scale patterns formed by solar active regions during the ascending phase of cycle 21}",
      journal = {\apj},
         year = 1983,
        month = feb,
       volume = {265},
        pages = {1056-1065},
          doi = {10.1086/160747},
       adsurl = {https://ui.adsabs.harvard.edu/abs/1983ApJ...265.1056G}
}

@ARTICLE{bumba1969,
       author = {{Bumba}, V. and {Howard}, R.},
        title = "{Solar Activity and Recurrences in Magnetic-Field Distribution}",
      journal = {\solphys},
         year = 1969,
        month = apr,
       volume = {7},
       number = {1},
        pages = {28-38},
          doi = {10.1007/BF00148402},
       adsurl = {https://ui.adsabs.harvard.edu/abs/1969SoPh....7...28B}
}

@ARTICLE{Fan2009,
       author = {{Fan}, Y.},
        title = "{Magnetic Fields in the Solar Convection Zone}",
      journal = {Living Reviews in Solar Physics},
         year = 2009,
        month = dec,
       volume = {6},
       number = {1},
        pages = {4},
          doi = {10.12942/lrsp-2009-4},
       adsurl = {https://ui.adsabs.harvard.edu/abs/2009LRSP....6....4F}
}

@ARTICLE{Cheung2010,
       author = {{Cheung}, M.~C.~M. and {Rempel}, M. and {Title}, A.~M. and {Sch{\"u}ssler}, M.},
        title = "{Simulation of the Formation of a Solar Active Region}",
      journal = {\apj},
         year = 2010,
        month = dec,
       volume = {720},
       number = {1},
        pages = {233-244},
          doi = {10.1088/0004-637X/720/1/233},
       adsurl = {https://ui.adsabs.harvard.edu/abs/2010ApJ...720..233C}
}

@ARTICLE{Castenmilleretal1986,
       author = {{Castenmiller}, M.~J.~M. and {Zwaan}, C. and {van der Zalm}, E.~B.~J.},
        title = "{Sunspot Nests - Manifestations of Sequences in Magnetic Activity}",
      journal = {\solphys},
         year = 1986,
        month = jun,
       volume = {105},
       number = {2},
        pages = {237-255},
          doi = {10.1007/BF00172045},
       adsurl = {https://ui.adsabs.harvard.edu/abs/1986SoPh..105..237C}
}

@misc{SILSO2026,
  author       = {{WDC-SILSO, Royal Observatory of Belgium}},
  title        = {Monthly Hemispheric Sunspot Numbers, Version 2.0},
  year         = {2025},
  howpublished = {\url{https://www.sidc.be/SILSO/datafiles}, World Data Center for the Sunspot Index and Long-term Solar Observations}
}
\end{document}